\documentclass[pre,superscriptaddress]{revtex4}
\usepackage{graphicx,amsmath,amsfonts}
\begin{document}
\title{On the first Sonine correction for granular gases}
\author{Fran\c{c}ois Coppex}
\affiliation{Department of Physics, University of Gen\`eve, 
CH-1211 Gen\`eve 4, Switzerland}
\author{Michel Droz}
\affiliation{Department of Physics, University of Gen\`eve, 
CH-1211 Gen\`eve 4, Switzerland}
\author{Jaros\l aw Piasecki}
\affiliation{Institute of Theoretical Physics, University of 
Warsaw, PL-00 681 Warsaw, Poland}
\author{Emmanuel Trizac}
\affiliation{Laboratoire de Physique Th\'eorique (UMR 8627 du Cnrs), B\^atiment 210, 
Universit\'e de Paris-Sud, 91405 Orsay, France}
\pacs{05.20.Dd, 73.23.Ad}
\begin{abstract}
We consider the velocity distribution for a granular gas of inelastic hard
spheres described by the Boltzmann equation. We investigate both 
the free of forcing case and a system heated by a stochastic force. 
We propose a new method to compute the first correction to Gaussian behavior
in a Sonine polynomial expansion quantified by the fourth cumulant $a_2$.
Our expressions are compared to previous results and to those obtained 
through the numerical solution of the Boltzmann equation. It is numerically shown that our method yields very accurate results for small velocities of the rescaled distribution. We finally discuss 
the ambiguities inherent to a linear approximation method in $a_2$. 
\end{abstract}
\maketitle

Most theories of rapid granular flows consider a granular 
gas as an assembly of inelastic hard spheres and
assume uncorrelated binary collisions
described by the Boltzmann equation, with a possible Enskog correction
to account for excluded volume effects 
\cite{McNamara,Brey1,Sela,NoijeErnst,Brey,Pre,Dufty,Montanero,Brilliantov,Cafiero,SM,Barrat}. 
The deviations from the Maxwellian velocity distribution may be accounted
for by an expansion in Sonine polynomials, 
and it is often sufficient to retain only
the leading term in this expansion, quantified by $a_2$, 
the  fourth cumulant of the velocity distribution 
\cite{Brey1,Montanero,Brey,Barrat1,Trizac3}. The purpose 
of this paper is twofold: first, we present a novel route 
to compute $a_2$, directly inspired from a method
that has been recently proposed to compute with accuracy 
the decay exponents
and non Maxwellian features of gas subjected to ballistic 
annihilation dynamics \cite{Trizac1,Trizac2} (where particles 
undergoing free flight motion disappear
upon contact \cite{BenNaim,Krapivsky}). In essence, this 
method considers the limit of vanishing velocities of the
Boltzmann equation, and
deduces $a_2$ from moments of the velocity distribution that are
{\it a priori} of lower order than those involved in the standard
derivation \cite{NoijeErnst,Montanero}.
We may consequently expect 
a better precision from this alternative approach, that is
analytically simpler to work out. We also know that the velocity
distribution is non Gaussian at high energies \cite{NoijeErnst,Montanero},
so that extracting the relevant kinetic information from the
behavior at vanishing velocities seems a promising route.
The second goal of this article is to discuss the ambiguities 
--common to both approaches-- encountered
performing computations up to linear order in $a_2$, neglecting
not only higher order Sonine contributions but also terms in $a_2^k$, $k=2,3$.
Such an ambiguity has first been mentioned by Montanero and Santos
\cite{Montanero}.

Within the framework of the Boltzmann equation, 
the one-particle velocity distribution function 
$f(\mathbf{v};t)$ for 
a homogeneous system free of forcing obeys the relation
\begin{equation}
\partial_t f(\mathbf{v}_1;t) = I(f,f), 
\label{eq1}
\end{equation}
where the collision integral reads
\begin{equation}
I(f,f) = \sigma^{d-1} \int_{\mathbb{R}^d} \mathrm{d}\mathbf{v}_2 
\int \mathrm{d}\widehat{\boldsymbol{\sigma}} \, 
\theta(\widehat{\boldsymbol{\sigma}} \cdot \widehat{\mathbf{v}}_{12}) 
(\widehat{\boldsymbol{\sigma}} \cdot \mathbf{v}_{12}) \left[ \frac{1}{\alpha^2} 
f(\mathbf{v}_1^{**};t) f(\mathbf{v}_2^{**};t) - f(\mathbf{v}_1;t)f(\mathbf{v}_2;t) 
\right]. 
\label{eq2}
\end{equation}
In Eq.~(\ref{eq2}), $\sigma$ is the diameter of the particles, $\theta$ the 
Heaviside distribution, $\mathbf{v}_{12} = \mathbf{v}_1 - \mathbf{v}_2$ 
the relative velocity of two particles, $\widehat{\mathbf{v}}_{12} = 
\mathbf{v}_{12}/v_{12}$, $v_{12} = |\mathbf{v}_{12}|$, and 
$\widehat{\boldsymbol{\sigma}}$ a unit vector joining
the centers of the grains. The space dimension is $d$.
The precollisional velocities $\mathbf{v}_i^{**}$  
and the postcollisional ones $\mathbf{v}_i$ are related by
\begin{subequations}
\label{eq3}
\begin{eqnarray}
\mathbf{v}_1^{**} &=& \mathbf{v}_1 - \frac{1+\alpha}{2 \alpha}(\mathbf{v}_{12} 
\cdot \widehat{\boldsymbol{\sigma}})\widehat{\boldsymbol{\sigma}}, \label{eq3a} \\
\mathbf{v}_2^{**} &=& \mathbf{v}_2 + \frac{1+\alpha}{2 \alpha}(\mathbf{v}_{12} 
\cdot \widehat{\boldsymbol{\sigma}})\widehat{\boldsymbol{\sigma}}, 
\label{eq3b}
\end{eqnarray}
\end{subequations} 
with $\alpha \in [0,1]$ the restitution coefficient. 
If energy is supplied to the system, an additional forcing term 
is present in Eq. (\ref{eq1}) \cite{Montanero}, but the general arguments
and method presented below remain valid. To be more specific, we shall
also consider the situation where the system is driven into a non equilibrium
steady state by a random force acting on the particles 
\cite{NoijeErnst,Pre,Montanero}. With this energy 
feeding mechanism, coined ``stochastic thermostat'', the Fokker-Planck
term $\xi_0^2 \nabla_{\mathbf{v}}^2 f$ should be added to the right-hand side (r.h.s.)
of Eq. (\ref{eq1}) \cite{NoijeErnst}, where $\xi_0$ is related to the amplitude
of the random force acting on the grains.

We are searching for an isotropic 
scaling solution  $\widetilde{f}(c)$ of Eq. (\ref{eq2}). The requirement
of a time independent behavior with respect to the typical
velocity $v_0(t) = \sqrt{2\langle \mathbf{v}^2 \rangle/d}$ imposes that
\cite{Brey1,NoijeErnst,Montanero}
\begin{equation}
f(\mathbf{v};t) = \frac{n}{v_0^d(t)}\widetilde{f}(c), 
\label{eq4}
\end{equation}
where the rescaled velocity is given by $c=v/v_0(t)$
and the angular brackets
$\langle \cdot \rangle$ denote the average over $f(\mathbf{v};t)$.
The presence of the density $n$ on the r.h.s. of Eq. 
(\ref{eq4}) ensures that $\int dc \widetilde{f}(c) = 1$ and 
$\int dc \, c^2 \widetilde{f}(c)=d/2$. 
This scaling function describing the homogeneous cooling state
satisfies the time-independent equation \cite{Brey1,NoijeErnst,Montanero}
\begin{equation}
\frac{\mu_2}{d}\left( d + c_1 \frac{\mathrm{d}}{\mathrm{d} c_1} \right) 
\widetilde{f}(c_1) = \widetilde{I}(\widetilde{f},\widetilde{f}), 
\label{eqlim1}
\end{equation}
where
\begin{equation}
\mu_p = - \int_{\mathbb{R}^d} d\mathbf{c}_1 \, c_1^p 
\widetilde{I}(\widetilde{f},\widetilde{f}), \label{eq5b}
\end{equation}
and
\begin{equation}
\widetilde{I}(\widetilde{f},\widetilde{f}) = 
\int_{\mathbb{R}^d} d\mathbf{c}_2 \int 
d\widehat{\boldsymbol{\sigma}} \, \theta(\widehat{\boldsymbol{\sigma}} 
\cdot \widehat{\mathbf{c}}_{12}) (\widehat{\boldsymbol{\sigma}} \cdot 
\mathbf{c}_{12}) \left[ \frac{1}{\alpha^2} f(c_1^{**}) 
f(c_2^{**}) - f(c_1) f(c_2) \right]. 
\label{eq6}
\end{equation}
It is useful to consider the hierarchy of moment  equations obtained by
integrating Eq.~(\ref{eqlim1}) over $c_1$ with 
weight $c_1^p$ \cite{NoijeErnst}
\begin{equation}
\mu_p = \frac{\mu_2}{d} p \langle c^p \rangle. 
\label{eq5}
\end{equation}

The solution of Eq. (\ref{eqlim1}) is non-Gaussian in several respects.
The high energy tail is overpopulated compared to the
Maxwellian \cite{NoijeErnst}, a generic although not systematic
feature for granular gases 
(a particular heating mechanism leading to an under-population 
at large velocities has been studied in \cite{Montanero}).
Deviation from Gaussian behavior may also be observed at thermal
scale or near the velocity origin. To study the latter correction, 
it is convenient to resort to a Sonine expansion
for the distribution function $\widetilde{f}(c)$ \cite{Landau}
\begin{equation}
\widetilde{f}(c) = \mathcal{M}(c) \bigg[ 1+ \sum_{i\geq 1}a_i S_i(c^2)\bigg], 
\label{eq7}
\end{equation}
where $\mathcal{M}(c) = \pi^{-d/2} \exp(-c^2)$ is the Maxwellian, and $S_i(c^2)$ 
the Sonine polynomials (that may be found in \cite{Landau}; the first few are
recalled in \cite{NoijeErnst}). 
Due to the constraint $\langle c^2\rangle = d/2$ the first correction $a_1$ vanishes \cite{NoijeErnst}, and for our purposes it is sufficient to know $S_2(x) = x^2/2 - (d+2)x/2+ d(d+2)/8$.
From Eq.~(\ref{eq7}) and making use of the orthogonality 
of the Sonine polynomials with respect to a Gaussian measure, one may
relate the coefficient $a_2$ to the kurtosis of the velocity distribution
\begin{equation}
\langle c^4 \rangle = \frac{d(d+2)}{4} (a_2 + 1), 
\label{eq9}
\end{equation}
so that, upon taking $p=4$ in Eq.~(\ref{eq5}), we get
\begin{equation}
\mu_4 = (d+2)(1+a_2) \mu_2. 
\label{eq10}
\end{equation}

\begin{figure}
\begin{center}
\includegraphics[width=0.6\columnwidth]{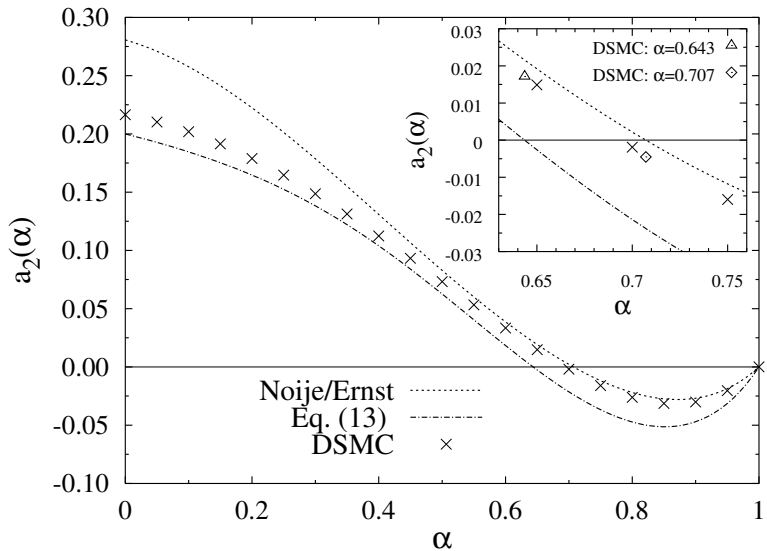}
\end{center}
\caption{Comparison of the correction $a_2(\alpha)$ for the free cooling 
in two dimensions obtained in \cite{NoijeErnst}, with Eq.~(\ref{eqlim3}). The crosses 
correspond to the ``exact'' result, obtained by solving the Boltzmann
equation with the DSMC method, for $10^6$ particles and 
approximately $500$ collisions for each particle. 
The inset is a zoom in the region of the smallest root of the fourth
cumulant.} 
\label{fig1}
\end{figure}

In the following analysis, we will only retain the first correction in the 
expansion (\ref{eq7}): $\widetilde f = \mathcal{M}(1+ a_2 S_2)$.
Computing $\mu_2$ and $\mu_4$ to linear order in $a_2$ 
with this functional ansatz [and further linearizing Eq. (\ref{eq10})], 
one deduces $a_2$ 
\cite{NoijeErnst,Montanero}. 
This approach is nonperturbative in the restitution coefficient. 
However, since the high energy tail of 
$\mathcal{M}(1+ a_2 S_2)$ is very distinct from that of the exact 
solution of Eq. (\ref{eqlim1}), computing $a_2$ from relation
(\ref{eq5}) with $p>4$ is expected to give a poor estimate,
all the worse as $p$ increases. With this in mind, it appears that
the limit of vanishing velocity of the rescaled Boltzmann equation
(\ref{eqlim1}) contains an interesting piece of 
information: 
\begin{equation}
\mu_2 \,\widetilde{f}(0) = \lim_{c_1 \to 0} 
\widetilde{I}(\widetilde{f},\widetilde{f}). 
\label{eqlim2}
\end{equation}
The main steps to compute this limit are given in appendix. Up to a geometrical prefactor, the loss term of $\lim \widetilde I$ 
on the r.h.s. reads $\widetilde f(0) \langle c_1\rangle$ and is thus of lower
order than the quantities appearing in (\ref{eq10}). 
Working at linear order in $a_2$, one may therefore expect to achieve 
a better accuracy when computing the various terms (except may be the gain term)
appearing in (\ref{eqlim2})
than in (\ref{eq10}).
In the context 
of ballistic annihilation, a related remark lead to analytical predictions
for the decay exponents of the dynamics and non-Gaussian features 
of the velocity statistics, in excellent agreement with the numerical
simulations \cite{Trizac1,Trizac2}. In the present situation, the 
gain term of $\widetilde{I}$ in (\ref{eqlim2}) cannot be written
as a collisional moment, so that the situation is less clear
and deserves some investigation. We propose to compare the value of $a_2$
following this route to the standard one of Refs. 
\cite{NoijeErnst,Montanero,Barrat1}. Evaluating
(\ref{eqlim2}) at first order in $a_2$, we obtain:
\begin{equation}
a_2 = \frac{4(\alpha^2+1)^2(\alpha^2-1)\left[ \sqrt{2}(\alpha^2+1)-2 \right] }{A(\alpha,d)}. 
\label{eqlim3}
\end{equation}
where
\begin{multline}
A(\alpha,d) = 5 + d(2-d) + 8\alpha (\alpha^2+1)(d-1) - \alpha^2(23-6d+d^2)+
\alpha^4(3+6d+d^2) \\
+\alpha^6(-1+2d+d^2)-\sqrt{2}(\alpha^2+1)^3(\alpha^2-1)(3+4d+2d^2)/4. 
\label{eqlim3b}
\end{multline}
In Fig.~\ref{fig1}, we compare this result with the analytical
expression of van Noije and Ernst \cite{NoijeErnst}. We also display the
fourth cumulant $a_2$ obtained by Monte Carlo simulations
from the numerical solution of the
nonlinear Boltzmann equation (\ref{eq1}) 
(so called DSMC technique \cite{Bird,DSMC}). Our expression 
appears more accurate at small inelasticity, but less satisfying 
close to elastic behavior. The smallest 
root of $a_2 = 0$ obtained with Eq.~(\ref{eqlim3}) is 
$\alpha^* = (\sqrt{2}-1)^{1/2} \simeq 0.643\ldots$. This root differs from 
the value $\alpha^{**} = 1/\sqrt{2} \simeq 0.707\ldots$ obtained upon 
solving~(\ref{eq10}) (both $\alpha^*$ and $\alpha^{**}$ do not depend
on space dimension $d$). The inset shows that the exact root is located in 
the interval $]\alpha^*,\alpha^{**}[$, and seems closer to
$\alpha^{**}$. 

\begin{figure}
\begin{center}
\includegraphics[width=0.6\columnwidth]{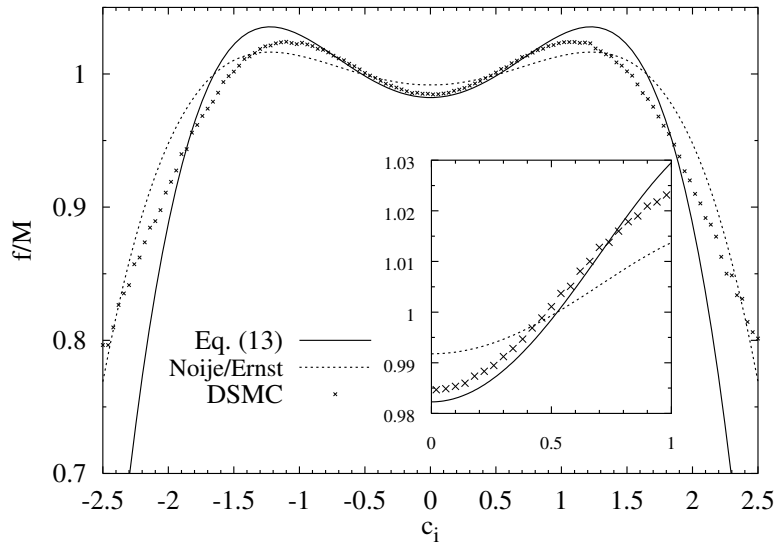}
\end{center}
\caption{Plot of $\widetilde{f}(c_i)/\mathcal{M}(c_i)$ for $\alpha = 0.8$. The curve labelled ``Eq.~(13)'' and ``Noije/Ernst'' correspond to $1+a_2 S_2$ where $a_2$ is given respectively by Eq.~(\ref{eqlim3}) and by the Sonine correction obtained by Noije and Ernst following the traditional route~\cite{NoijeErnst}. ``DSMC'' refers to the full distribution obtained from the solution of the Boltzmann equation (using $10^6$ particles and averaging over $300$ independent samples).}
\label{new1} 
\end{figure}

\begin{figure}
\begin{center}
\includegraphics[width=0.6\columnwidth]{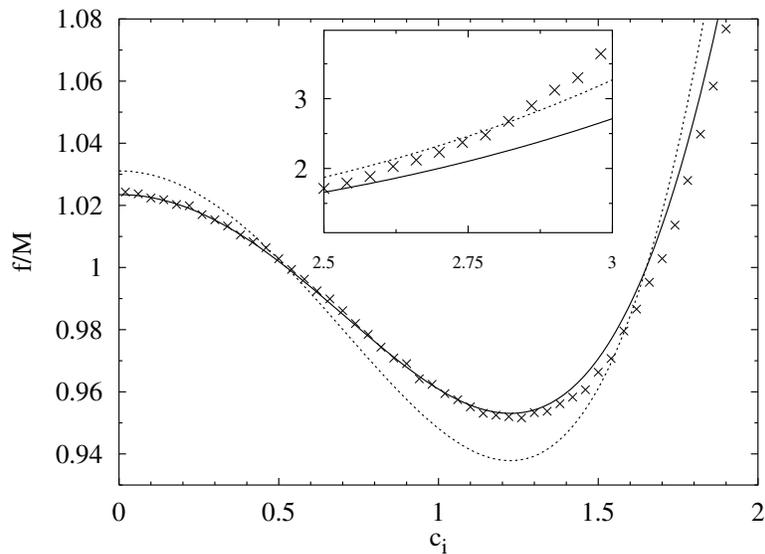}
\end{center}
\caption{Same as Fig.~\ref{new1} for $\alpha=0.5$.}
\label{new2} 
\end{figure}

In order to understand the discrepancy close to the elastic limit shown in Fig.~\ref{fig1}, it is useful to study the first Sonine correction $\widetilde{f}(c_i)/\mathcal{M}(c_i) = 1 + a_2 S_2(c_i^2)$. The result for $\alpha = 0.8$ where our method seems to be the less accurate is shown in Fig.~\ref{new1}, and in Fig.~\ref{new2} for $\alpha=0.5$.

\begin{figure}
\begin{center}
\includegraphics[width=0.6\columnwidth]{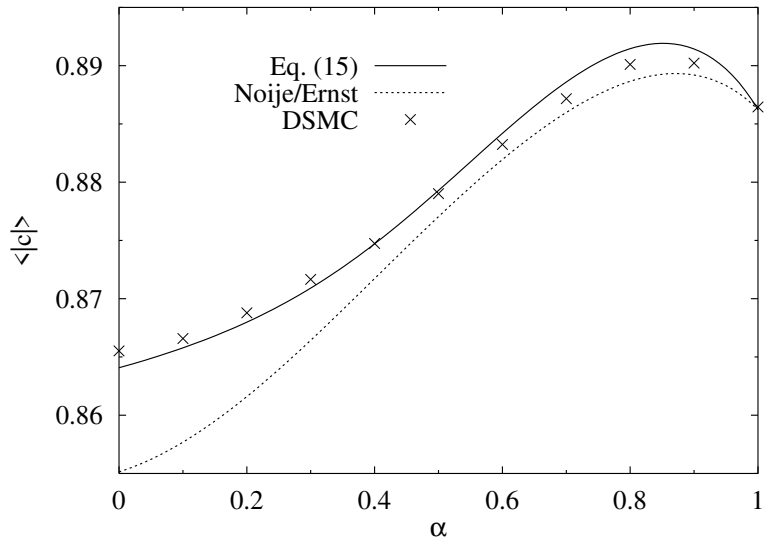}
\end{center}
\caption{First rescaled velocity moment $\langle |c| \rangle$ as a function of the restitution coefficient. DSMC is done for $10^5$ particles and approximately 500 collisions for each particle.}
\label{new3}
\end{figure}

In spite of the imprecision of our analytical expression for $a_2$ seen in Fig.~\ref{fig1}, Fig.~\ref{new1} shows that the limit method is very accurate for small velocities, but turns to quickly become more imprecise for bigger velocities. This suggests that computing the fourth cumulant from 
the limit of vanishing velocities gives more weight to this
region which leads to a better behavior of the Sonine
expansion for small velocities. On the other hand, the traditional route yields a global interpolation for all velocities. The good precision of our result for small velocities and the lower accuracy for higher velocities is confirmed in Fig.~\ref{new2}. Exploiting the above qualitative interpretation of the limit method, we expect to archieve a good accuracy using Eq.~(\ref{eqlim3}) in order to find the first moment~\cite{Trizac2}:
\begin{equation}
\langle |c| \rangle = \frac{\sqrt{\pi}}{2} \left( 1 - \frac{a_2}{8} \right). \label{star}
\end{equation}
Indeed, we suppose that the function $a_2$ obtained from the limit method gives a precise description of the rescaled velocity distribution for small velocities. Thus our $a_2$ is likely to describe more accurately a low order velocity moment than a high order one. This is confirmed by Fig.~\ref{new3}.

\begin{figure}
\begin{center}
\includegraphics[width=0.6\columnwidth]{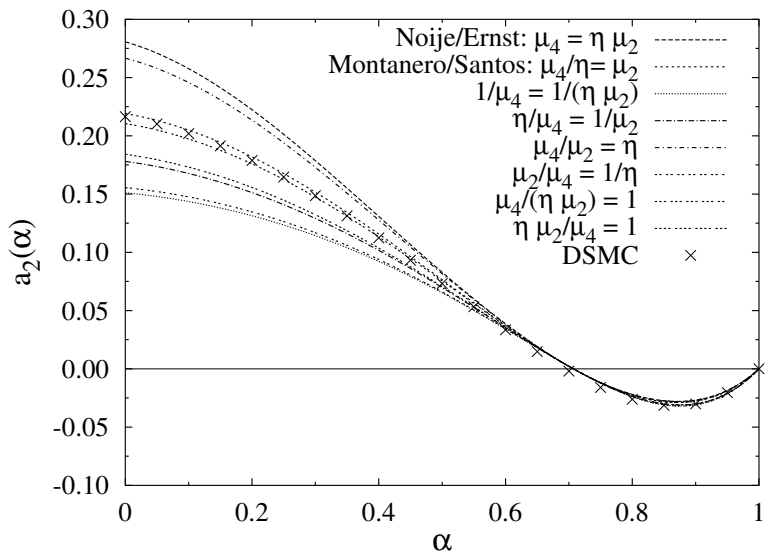}
\end{center}
\caption{The eight possible fourth cumulant $a_2$ obtained 
from Eq. (\ref{eq10}), corresponding to the two-dimensional
homogeneous free cooling. 
We define $\eta = (d+2)(1+a_2)$, 
then rewrite the equation 
$\mu_4 = \eta \mu_2$ according to the eight possible different combinations
mentioned in the legend,  
before doing the linear Taylor expansion around $a_2 = 0$. The first curve 
is the plot of the function $a_2$ obtained by 
van Noije and Ernst \cite{NoijeErnst}, whereas the second one -- 
obtained by Montanero and Santos~\cite{Montanero}-- is very close
to the exact results shown by crosses.} 
\label{fig2} 
\end{figure}

\begin{figure}
\begin{center}
\includegraphics[width=0.6\columnwidth]{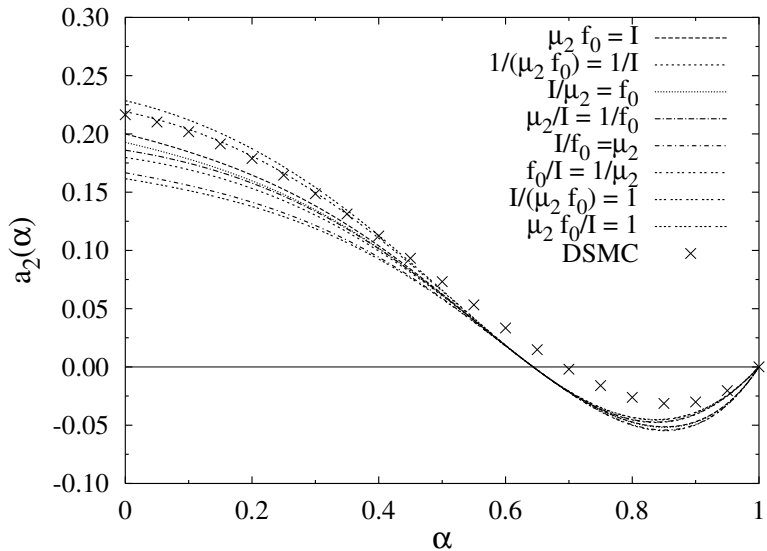}
\end{center}
\caption{Same as Fig. \ref{fig2}, making use of Eq. (\ref{eqlim2}) instead of
(\ref{eq10}) to compute the first Sonine correction. In the legend, $I$ denotes
$\lim \widetilde I$ and $f_0 = \widetilde f(0)$.} 
\label{fig3} 
\end{figure}

As emphasized by Montanero and Santos \cite{Montanero}, a certain
degree of ambiguity is present when evaluating an identity
such as (\ref{eq10}) or (\ref{eqlim2}) to first order in $a_2$. 
According to the way we rearrange the terms $\mu_4$, $\mu_2$, 
and $(d+2)(1+a_2)$ in say Eq.~(\ref{eq10}) and subsequently 
apply a Taylor series expansion 
in $a_2$, we obtain different predictions for $a_2(\alpha)$. 
For instance van Noije and Ernst
did expand the relation~(\ref{eq10})~\cite{NoijeErnst}, whereas 
Montanero and Santos also considered other possibilities such as
$\mu_4/\mu_2 = (d+2)(1+a_2)$ (this leads to a  
result which turns out to be fairly close to the one in \cite{NoijeErnst}) and 
also $\mu_4/(1+a_2) = (d+2) \mu_2$. For small $\alpha$ in the latter case,  
the resulting $a_2$ turns out to be
20$\%$ lower than the previous ones,
and very close to the exact (within Boltzmann's equation framework) 
numerical results, for all the values of the
restitution coefficient \cite{Montanero}.
We push further this remark and show in 
Fig.~\ref{fig2} the eight simplest different possible 
functions $a_2(\alpha)$ obtained upon rearranging the terms of Eq.~(\ref{eq10}) 
and expanding the result to first order in $a_2$.
A similar ambiguity is present making use of Eq. (\ref{eqlim2}).
The corresponding eight different possibilities are plotted in Fig. \ref{fig3}.
It appears that the envelope of the curves following from this method
is less spread than within the ``traditional'' route, by a factor of 
approximately 2. We thus achieve a better accuracy at small $\alpha$.

The dispersion of the curves in Figs. \ref{fig2} and \ref{fig3} illustrates
the nonvalidity of the linearization approximation at small $\alpha$.
However --and concentrating on Fig. \ref{fig2}-- it appears that
all curves do not have the same status. Brilliantov and P\"oschel
have indeed solved analytically the full nonlinear problem 
[i.e. working again with the distribution function 
$\widetilde f = \mathcal{M}(1+ a_2 S_2)$ but keeping nonlinear terms
in $a_2$], and obtained results that are very close to those of 
Noije/Ernst, except for $\alpha<0.2$ where they found slightly 
larger fourth cumulants \cite{Brilliantov}. Their result is therefore
farther away from the exact one obtained by DSMC (see e.g. Fig.
\ref{fig1} where it appears than the Noije/Ernst expression already 
overestimates the exact curve). The difference between the DSMC
results and those of Brilliantov/P\"oschel therefore illustrates
the relevance of Sonine terms $a_i$ with $i \geq 3$
in expansion (\ref{eq7}). However, some of the curves shown in 
Fig. \ref{fig2} lie close to the exact one, which means 
that it is possible to correct the deficiencies of truncating
$\widetilde f$ at second Sonine order by an ad-hoc linearizing scheme.
The agreement obtained is nevertheless incidental, and the 
corresponding analytical expression should be considered as a semi-empirical
interpolation supported by numerical simulations. One should thus emphasize that the right way to compute $a_2$ is to use its definition involving the fourth rescaled velocity cumulant of Eq.~(\ref{eq9}) because this relation is not sensitive to higher order Sonine terms, nor to nonlinearities, even if this route doesn't give the most accurate description in the small velocity domain (as seen from Figs.~\ref{new1} and~\ref{new2})

\begin{figure}
\begin{center}
\includegraphics[width=0.6\columnwidth]{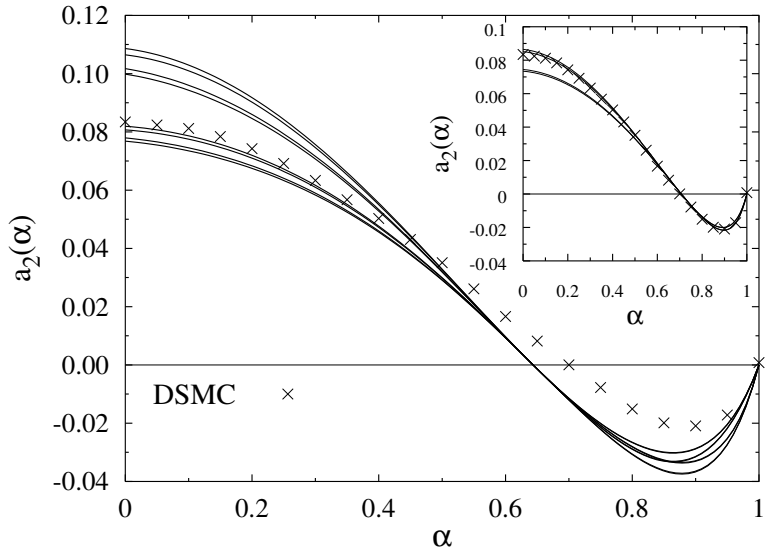}
\end{center}
\caption{The counterpart of Fig. \ref{fig3} for the two dimensional 
stochastic thermostat. The inset shows the 4 possibilities associated with 
the method of Refs. \cite{NoijeErnst,Montanero}. The symbols
show the results of DSMC simulations.
} 
\label{fig4} 
\end{figure}

For completeness, we now briefly consider the stochastic thermostat situation
\cite{NoijeErnst,Pre,Montanero,Trizac3}, where the counterpart of
Eq. (\ref{eqlim1}) reads
\begin{equation}
-\frac{\mu_2}{2d} \,\nabla^2_{\mathbf{c}_1} \widetilde f(c_1) \, = \, \widetilde I
(\widetilde f,\widetilde f).
\label{eqsto}
\end{equation}
Considering again the limit $c_1 \to 0$ and retaining only the first correction in the expansion (\ref{eq7}), we get
\begin{equation}
\frac{\mu_2}{2 \pi^{d/2}}\,  \left[ 2 + a_2 \frac{(d+2)(d+4)}{4} \right] \,=\,
\lim_{c_1\to 0}
\widetilde I
(\widetilde f,\widetilde f).
\label{eqlimsto}
\end{equation}
Given that the r.h.s. is already known from the free cooling calculation,
it is straightforward to extend the previous results to the present case. 
As before, there are 8 possible ways to extract $a_2$ from  Eq. 
(\ref{eqlimsto}) working at linear order. The resulting expressions 
are displayed in  Fig. \ref{fig4}. On the other hand, the moment method 
described in Refs. \cite{NoijeErnst,Montanero} makes use of the identity
$\mu_2 (d+2) = \mu_4$, that is a direct consequence of Eq. (\ref{eqsto}).
There are thus 4 possible rearrangements leading to the different 
cumulants shown in the inset of Fig. \ref{fig4}. For comparison, we have also
implemented Monte Carlo simulations in the present heated situation
(see the crosses in Fig. \ref{fig4}). It is difficult to compare the
dispersion of the curves with both methods (8 possibilities versus 4), 
since our approach makes
use of Eq.~(\ref{eqlimsto}) which is of higher order in $a_2$ than 
$\mu_2 (d+2) = \mu_4$,
the starting point used in Refs. \cite{NoijeErnst,Montanero}.
Our method appears here less accurate than for the free cooling, with again
an underestimation of $a_2$ at large $\alpha$.

\begin{figure}
\begin{center}
\includegraphics[width=0.6\columnwidth]{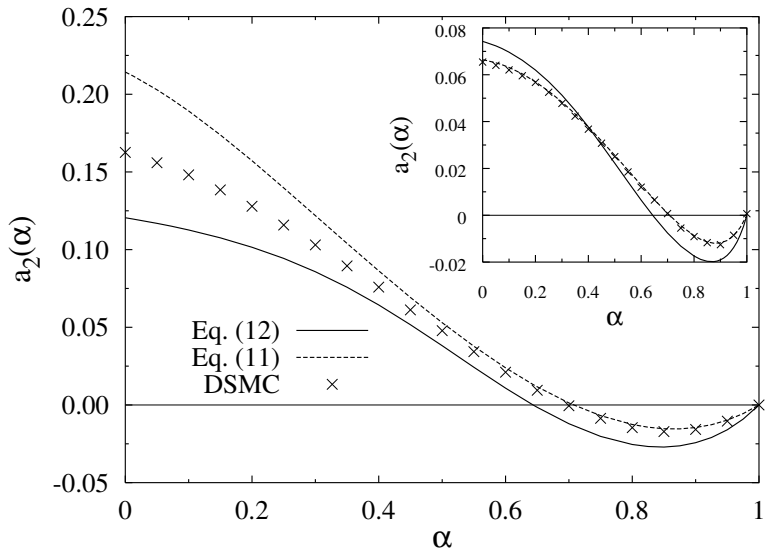}
\end{center}
\caption{Fourth cumulant in $3$ dimensions for a force free system in the
regime of homogeneous cooling. The curves correspond to the nonlinear
solutions of Eqs. (\ref{eq10}) and (\ref{eqlim2}) (see text for details).
The crosses correspond to the Monte Carlo results.
The inset shows the same curves for the 
stochastic thermostat.} 
\label{fig5} 
\end{figure}

In order to get free from the ambiguities inherent to a linear computation in $a_2$,
we have also solved the full nonlinear problem. The computation becomes
cumbersome, and since Brilliantov and P\"oschel \cite{Brilliantov} 
have already initiated this
route in 3D for the homogeneous free cooling 
(thereby providing the calculation of $\mu_2$ and $\mu_4$), we will 
turn our attention to the 3D situation. First and for the
sake of comparison, we have repeated the nonlinear
derivation of Ref. \cite{Brilliantov} for the stochastic thermostat.
Second, we have computed the right-hand sides of Eqs. (\ref{eqlim2}) and 
(\ref{eqlimsto}) without any linearization,
from the form $\widetilde f = \mathcal{M}(1+ a_2 S_2)$. The left-hand
sides only require the knowledge of $\mu_2$. For both free and forced
situations, we subsequently obtain a polynomial 
equation of degree 3 for $a_2$ from which we extract the physical root,
the two others corresponding to unstable scaling solutions
\cite{Brilliantov}. The results are displayed in Fig. \ref{fig5}.
In particular, our approach again suffers from an underestimation
of $a_2$ for $\alpha>0.5$, already observed within the linear
computation, and that is thus ascribable to 
Sonine terms of order 3 or higher. In this respect, it is surprising
that these terms do not affect similarly the moment
method of Ref. \cite{Brilliantov} in the same range of inelasticities.

To sum up, using a new approach we obtain the first non-Gaussian 
correction $a_2$ to the scaled velocity distribution. 
%At small inelasticity and working at linear order in $a_2$,
%this correction turns out to be closer to the exact results obtained numerically 
%than its counterpart obtained by van Noije and Ernst~\cite{NoijeErnst}. 
%On the other hand, close to the elastic limit, the latter expression is superior
%to that derived here. 
In view of the above results, 
we conclude that our approach constitutes an improvement over the previous procedures in the small velocity regime, and our analysis turns to be technically simpler to perform.
We have also discussed the ambiguities
that arise 1) when restricting ourselves to second Sonine
order, and 2) when a further linearization 
of the various relevant relations is performed. It appears that
an  {\it ad-hoc} linearization scheme (point 2) may circumvent the 
limitations inherent to point 1. 
In any case, the computation of a non-Gaussian correction suffers
from uncontrolled approximations that systematically need to 
be confronted against 
numerical simulations.

%\begin{acknowledgments} 
We acknowledge useful discussions with A. Santos and A. Barrat.
This work was partially supported by the Swiss National Science Foundation and 
the French ``Centre National de la Recherche Scientifique''.
%\end{acknowledgments}

%==================================================================
\appendix*
\section{Calculation of the limit $c_1 \to 0$ of the collision term}
We define the loss term $\widetilde{I}_l$ and gain term $\widetilde{I}_g$ by
\begin{subequations}
\label{appi}
\begin{eqnarray}
\widetilde{I}_l &=& - \lim_{c_1 \to 0} \int_{\mathbb{R}^d} d\mathbf{c}_2 \int d\widehat{\boldsymbol{\sigma}} \, \theta(\widehat{\boldsymbol{\sigma}} \cdot \widehat{\mathbf{c}}_{12}) (\widehat{\boldsymbol{\sigma}} \cdot \mathbf{c}_{12}) \widetilde{f}(c_1)\widetilde{f}(c_2)\label{loss}\\
\widetilde{I}_g &=& \lim_{c_1 \to 0} \frac{1}{\alpha^2} \int_{\mathbb{R}^d} d\mathbf{c}_2 \int d\widehat{\boldsymbol{\sigma}} \, \theta(\widehat{\boldsymbol{\sigma}} \cdot \widehat{\mathbf{c}}_{12}) (\widehat{\boldsymbol{\sigma}} \cdot \mathbf{c}_{12}) \widetilde{f}(c_1^{**}) \widetilde{f}(c_2^{**}),\label{gain} 
\end{eqnarray}
\end{subequations}
so that $\lim_{c_1 \to 0} \widetilde{I}(\widetilde{f},\widetilde{f}) = \widetilde{I}_l + \widetilde{I}_g$.

Taking the limit $c_1 \to 0$ of the loss term yields the exact result
\begin{equation}
\widetilde{I}_l = - \beta_1 \widetilde{f}(0) \langle c_2 \rangle,\label{loss1}
\end{equation}
where
\begin{equation}
\beta_1 = \int_{\mathbb{R}^d} d\widehat{\boldsymbol{\sigma}} \, \theta(\widehat{\boldsymbol{\sigma}} \cdot \widehat{\mathbf{c}}_{2}) (\widehat{\boldsymbol{\sigma}} \cdot \widehat{\mathbf{c}}_{2}) = \frac{\pi^{(d-1)/2}}{\Gamma\left[(d+1)/2\right]},
\end{equation}
with $\Gamma$ the gamma function. Within the framework of the Sonine expansion (\ref{eq7}), neglecting the coefficients $a_i$, $i \geq 3$, and making use of \cite{Trizac2}
\begin{equation}
\langle c_2 \rangle = \left( 1- \frac{a_2}{8}\right) \frac{\Gamma\left[(d+1)/2\right]}{\Gamma(d/2)},
\end{equation}
Eq. (\ref{loss1}) becomes
\begin{equation}
\widetilde{I}_l = - \frac{S_d \mathcal{M}(0)}{2 \sqrt{\pi}} \left[ 1 + a_2 \frac{d(d+2)}{8} \right] \left( 1- \frac{a_2}{8}\right), \label{perte7}
\end{equation}
where $S_d = 2 \pi^{d/2}/\Gamma(d/2)$ is the surface of the $d$-dimensional sphere.

Defining $\beta = (1+\alpha)/(2\alpha) > 1$, the precollisional rescaled velocities $\mathbf{c}_i^{**}$ and postcollisional ones $\mathbf{c}_i$ are related by
\begin{subequations}
\label{eqc}
\begin{eqnarray}
\mathbf{c}_1^{**} &=& \mathbf{c}_1 - \beta(\mathbf{c}_{12} 
\cdot \widehat{\boldsymbol{\sigma}})\widehat{\boldsymbol{\sigma}}, \label{eqc1} \\
\mathbf{c}_2^{**} &=& \mathbf{c}_2 + \beta(\mathbf{c}_{12} 
\cdot \widehat{\boldsymbol{\sigma}})\widehat{\boldsymbol{\sigma}}.
\label{eqc2}
\end{eqnarray}
\end{subequations} 
The gain term (\ref{gain}) thus becomes
\begin{equation}
\widetilde{I}_g = \frac{1}{\alpha^2} \int_{\mathbb{R}^d} d\mathbf{c}_2 \int d\widehat{\boldsymbol{\sigma}} \, \theta(\widehat{\boldsymbol{\sigma}} \cdot \widehat{\mathbf{c}}_{2}) (\widehat{\boldsymbol{\sigma}} \cdot \mathbf{c}_{2}) \, \widetilde{f} \left[ \beta (\mathbf{c}_2 \cdot \widehat{\boldsymbol{\sigma}}) \widehat{\boldsymbol{\sigma}} \right] \widetilde{f} \left[ \mathbf{c}_2 - \beta (\mathbf{c}_2 \cdot \widehat{\boldsymbol{\sigma}}) \widehat{\boldsymbol{\sigma}}  \right], \label{gain2}
\end{equation}
where the function $\widetilde{f}$ is isotropic. Performing the integration over $\mathbf{c}_2$ before that over $\hat{\boldsymbol{\sigma}}$, we choose the $x$ Cartesian coordinate as corresponding to the $\hat{\boldsymbol{\sigma}}$ direction. The velocity $\mathbf{c}_2$ is thus written $\mathbf{c}_2 = c_x \widehat{\mathbf{x}} + \mathbf{c}_\perp$, with $c_x = (\mathbf{c}_2 \cdot \widehat{\boldsymbol{\sigma}}) \in \mathbb{R}$ and $\mathbf{c}_\perp = \mathbf{c}_2 - c_x \widehat{\mathbf{x}} \in \mathbb{R}^{d-1}$. Eq. (\ref{gain2}) becomes
\begin{eqnarray}
\widetilde{I}_g &=& \frac{1}{\alpha^2} \int d\widehat{\boldsymbol{\sigma}} \int_{\mathbb{R}^d} d\mathbf{c}_2 \, \theta(c_x) \, c_x \widetilde{f} (\beta c_x \widehat{\boldsymbol{\sigma}}) \widetilde{f} (\mathbf{c}_2 - \beta c_x \widehat{\boldsymbol{\sigma}}) \\
&=& \frac{S_d}{\alpha^2} \int_0^\infty dc_x \, c_x \int_{\mathbb{R}^{d-1}} d\mathbf{c}_\perp \, \widetilde{f} (\beta c_x) \widetilde{f} \left( \sqrt{c_\perp^2 + c_x^2(1-\beta)^2 } \right). \label{gain3}
\end{eqnarray}
Eq. (\ref{gain3}) is an exact relation within Boltzmann's framework. Making use of the the Sonine expansion (\ref{eq7}) where we retain only the first correction $a_2$, Eq. (\ref{gain3}) becomes
\begin{equation}
\widetilde{I}_g = \frac{S_d}{\alpha^2 \pi^d} \int_0^\infty dc_x \, c_x \mathrm{e}^{- \left[ \beta^2 + (1-\beta^2) \right]c_x^2} \int_{\mathbb{R}^{d-1}} d\mathbf{c}_\perp \, \mathrm{e}^{- c_\perp^2}\left[ 1+a_2 S_2(\beta^2 c_x^2) \right] \left\{ 1+a_2 S_2\left[c_\perp^2 + c_x^2 (1-\beta)^2\right] \right\}. \label{gain4}
\end{equation}
With the definition of the second Sonine polynomial $S_2(x) = x^2/2 - (d+2)x/2 + d(d+2)/8$, one sees that Eq. (\ref{gain4}) may be expressed as a sum of products of the integrals
\begin{subequations}
\label{eqj}
\begin{eqnarray}
J_\perp(n) &=& \int_{\mathbb{R}^{d-1}} d\mathbf{c}_\perp \, \mathrm{e}^{- c_\perp^2} c_\perp^n, \label{eqja} \\
J_x(n) &=& \int_0^\infty dc_x \, \mathrm{e}^{- \left[ \beta^2 + (1-\beta^2) \right]c_x^2} c_x^n, \label{eqjb}
\end{eqnarray}
\end{subequations} 
that may be computed using the general relation ($a>0$)
\begin{equation}
\int_{\mathbb{R}^d} d\mathbf{x} |\mathbf{x}|^n \mathrm{e}^{-a \mathbf{x}^2} = \frac{\pi^{d/2}}{a^{(d+n)/2}} \frac{\Gamma\left[(d+n)/2\right]}{\Gamma(d/2)}.
\end{equation}
Tedious but technically simple calculations thus lead to
\begin{equation}
\widetilde{I}_g = \frac{S_d \mathcal{M}(0)}{2 \sqrt{\pi}} \left[ \frac{2}{1+\alpha^2} + a_2 D_1(\alpha,d) + a_2^2 D_2(\alpha,d) \right],\label{gain7}
\end{equation}
where the final expressions $D_1(\alpha,d)$ and $D_2(\alpha,d)$ are too cumbersome to be given here. Finally, the limit $c_1 \to 0$ of Eq. (\ref{eq6}) is given by the sums of Eqs. (\ref{perte7}) and (\ref{gain7}).

%==================================================================


\begin{thebibliography}{}
\bibitem{McNamara} S. McNamara and W.R. Young, 
         Phys. Fluids A {\bf 5}, 34 (1993).
\bibitem{Brey1} J.J.~Brey, M.J. Ruiz-Montero, and D. Cubero, 
         Phys.~Rev.~E~\textbf{54}, 3664 (1996).
\bibitem{Sela} N. Sela and I. Goldhirsch, J. Fluid Mech. {\bf 361}, 41 (1998).
\bibitem{NoijeErnst} T.P.C.~van~Noije and M.H.~Ernst, 
         Granular Matter~\textbf{1}, 57 (1998) (e-print: cond-mat/9803042).
\bibitem{Brey} J.J.~Brey, J.W.~Dufty, C.S.~Kim, and A.~Santos, 
         Phys.~Rev.~E~\textbf{58}, 4638 (1998).  
\bibitem{Pre} T.P.C. van Noije, M.H. Ernst, E. Trizac, and I. Pagonabarraga, 
         Phys.\ Rev.\ E {\bf 59}, 4326 (1999).
\bibitem{Dufty} V.~Garz\'o and J.~Dufty, 
         Phys.~Rev.~E~\textbf{60}, 5706 (1999).  
\bibitem{Montanero} J.M.~Montanero and A.~Santos, 
         Granular~Matter~\textbf{2}, 53 (2000) (e-print: cond-mat/0002323). 
\bibitem{Brilliantov} N.V.~Brilliantov and T.~P\"oschel, 
         Phys.~Rev.~E~\textbf{61}, 2809 (2000).
\bibitem{Cafiero} R. Cafiero, S. Luding, and H.J. Herrmann, 
         Phys. Rev. Lett. {\bf 84}, 6014 (2000).
\bibitem{SM} R. Soto and M. Mar\'eschal, 
         Phys. Rev. E {\bf 63}, 041303 (2001).
\bibitem{Barrat} A. Barrat and E. Trizac, 
     Phys. Rev. E {\bf 66}, 051303 (2002). 
\bibitem{Trizac3} A.~Barrat, T.~Biben, Z.~R\'acz, E.~Trizac, and F.~van~Wijland, 
         J.~Phys.~A~\textbf{35}, 463 (2002) (e-print: cond-mat/0110345).
\bibitem{Barrat1} A. Barrat, E. Trizac, and J.N. Fuchs, 
         Eur. Phys. J. E {\bf 5}, 161 (2001).
\bibitem{Trizac1} E. Trizac, 
         Phys. Rev. Lett. {\bf 88}, 160601 (2002).   
\bibitem{Trizac2} J.~Piasecki, E.~Trizac, and M.~Droz, 
         Phys.~Rev.~E~\textbf{66}, 066111 (2002).
\bibitem{BenNaim}  E. Ben-Naim, S. Redner, and F. Leyvraz, 
         Phys. Rev. Lett. {\bf 70}, 1890 (1993). 
\bibitem{Krapivsky} P.L. Krapivsky and C. Sire, 
         Phys. Rev. Lett. {\bf 86}, 2494 (2001). 
\bibitem{Landau} L. Landau and E. Lifshitz, 
         {\it Physical Kinetics}, Pergamon Press (1981).     
\bibitem{Bird} G. Bird
         {\it Molecular Gas Dynamics and the Direct Simulation of Gas Flows},
         Clarendon Press, Oxford, (1994).
\bibitem{DSMC} J.M. Montanero, V. Garz\'o, A. Santos, and J.J. Brey,
         {\it J. Fluid Mech.} {\bf  389}, 391 (1999).
\end{thebibliography}
\end{document}